# A Study of the correlation between electrical resistivity and matric suction for unsaturated ash-fall pyroclastic soils in The Campania region (Southern Italy)

De Vita P., Di Maio R., Piegari E.*

*Department of Earth Sciences - University of Naples "Federico II"- Largo San Marcellino, 10 – 80132 Napoli, Italy*

**ABSTRACT**

In the territory of the Campania region (southern Italy), critical rainfall events periodically trigger dangerous fast slope movements involving ashy and pyroclastic soils originated by the explosive phases of the Mt. Somma-Vesuvius volcano and deposited along the surrounding mountain ranges. In this paper, an integration of engineering-geological and geophysical measurements is presented to characterize unsaturated pyroclastic samples collected in a test area on the Sarno Mountains (Salerno and Avellino provinces, Campania region). The laboratory analyses were aimed at defining both soil water retention and electrical resistivity curves versus water content. From the matching of the experimental data, a direct relationship between electrical resistivity and matric suction is retrieved for the investigated soil horizons typical of a ash-fall pyroclastic succession. The obtained relation turns out to be helpful in characterizing soils up to close saturation, which is a critical condition for the trigger of slope failure. In such a regime, the water content and the matric suction have small variations, while electrical resistivity variations can be appreciated in a larger range of values. For this reason, besides suction measurements on very small soil volumes through classical tensiometers, our analyses suggest the direct monitoring of in-situ electrical resistivity values as an effective tool to recognise the hydrological state of larger and more representative soil volumes and to improve early warning of dangerous slope movements.

## 1. INTRODUCTION

Since historical times, the instability of volcanoclastic soils that mantle the Campanian carbonatic ridges surrounding the Somma-Vesuvius volcano represents a significant natural hazard for the numerous underlying urbanized areas (de Riso and Nota d'Elogio, 1973; Celico et al., 1986; Guadagno, 1991). Landslides are basically due to the instability of ash-fall pyroclastic deposits covering sedimentary rocky steep slopes, which are often triggered by heavy and prolonged rainfalls, thus making these areas geologically fragile environment**s**. In addition, these phenomena were more recently recognised in other Campanian volcanic areas such as the Island of Ischia (Del Prete and Mele, 2006; De Vita et al., 2007; Di Maio et al., 2007) and the hilly zones around the city of Naples (Calcaterra et al., 2002).

*Correspondence to: E. Piegari (esterpiegari@gmail.com ; +39 081 2538377)

Special attention has been paid to rainfall-induced shallow landslides in the last ten years, following the calamitous event that occurred on 5th and 6th May 1998 in the Sarno Mountains. This caused the loss of 156 lives and serious damage in the towns located at the foothill of the Pizzo d'Alvano relief. Various aspects involved in the triggering mechanisms and dynamical evolution of these landslides were analysed. Many studies focused on recognising and mapping stratigraphic and morphological conditions for which landslide susceptibility results higher (Celico and Guadagno, 1998; Di Crescenzo and Santo, 2005; Guadagno et al., 2005; De Vita et al., 2006; Perriello Zampelli, 2009). Other studies have been focused on the understanding and modelling of hydrological processes characterising the pyroclastic soil mantle during extreme hydrological event as well as on the modelling of inherent stability, also considering shear strength of pyroclastic soils (Cascini et al., 2000; Crosta and Dal Negro, 2003; Frattini et al., 2004; Cascini et al., 2008).

Another investigated key aspect regarded landslide hazard analysis resulting from rainfall events. On the basis of the evaluation of intensity and rainfall amounts that provoked landslides, several empirical hydrological models were developed to determine hydrological thresholds above which landslides occurred and/or is very likely that they occur (Chirico et al., 2000; De Vita, 2000; Fiorillo and Wilson, 2004). These thresholds identify high intensity/duration condition of rainfall when compared with other hydrological thresholds determined worldwide (Baum and Godt, 2009; Guzzetti et al., 2008). A further approach was aimed to model hydrological processes induced by heavy rainfalls by which pore pressure distribution within pyroclastic mantle, both in unsaturated and saturated domains, was achieved (Basile et al., 2003; Crosta and Dal Negro, 2003; Sorbino, 2005).

More recently, to analyze the susceptibility to sliding of pyroclastic covers, an alternative approach has been proposed by two of us, where an integration of geophysical and statistical methods is attempted (Piegari et al., 2009a; Piegari et al. 2009b). In particular, the authors relate electrical resistivity to stability by introducing an empirical local safety factor that explicitly depends on in-situ measurements of electrical resistivity values. The time evolution of the proposed safety factor, which is caused by external perturbations, like rainfalls, is studied by means of a cellular automaton model (Piegari et al., 2006a; Piegari et al., 2006b) that allows locating areas that are more likely than others to be susceptible to slide. In this framework, therefore, the electrical characterization of pyroclastic soils as dependent on water content is fundamental due to the straight correlation between rainfalls and electrical resistivity variations. Furthermore, since soil matric suction is also strongly dependent on water content and, thus, a key factor for stability analysis of steep slopes covered by pyroclastic soils (e.g. Crosta and Dal Negro, 2003; Sorbino, 2005) as well as other soil-mantled slopes (Godt et al., 2009), we argue that a study of the possible relationship between suction and electrical resistivity might provide an useful contribution to an improved characterization of the conditions leading to rainfall-induced instability. We consider such relationships as useful tools for assessing both water content and suction more rapidly and in a soil volume larger than that investigated by punctual measurements obtained through tensiometers. These relationships are therefore also more representative of landslide triggering.

*Correspondence to: E. Piegari (esterpiegari@gmail.com ; +39 081 2538377)

In this paper, we report the results of the above mentioned study carried out on soil samples representative of three soil horizons typical of pyroclastic successions that cover most mountain slopes that surround Campanian volcanic centres. The analysis consisted in laboratory experiments finalised to study the behaviour of the soil matric suction and the electrical resistivity at varying of the soil water content as well as to retrieve empirical relationships for both hydraulic parameters and the geophysical one.

## 2. SOIL SAMPLE COLLECTING

Experimental laboratory analyses were carried out on pyroclastic soil specimens collected in two neighbouring sites in a sample area located on Mount Pizzo d'Alvano (1133 m a.s.l.), belonging to the Sarno Mountain Range (Fig. 1). The sampling sites are located above the initial detachment areas of two landslides that occurred on the 5$^{th}$ and 6$^{th}$ of May 1998 (Fig. 2a). After the initial instability, they spread on the slope as avalanches and then flowed in a very steep hydrographic channel, known as Vallone Tuostolo, and finally struck the town of Episcopio (80 m a.s.l.) located at its outlet. The sample areas are situated at 725 m a.s.l. in a sector of slope whose longitudinal profile is convex, presenting a progressive increase in slope angle from 15° up to verticality. Pyroclastic soil samples were collected in proximity of the main scarps of the initial landslides characterised by a slope angle of about 40° and by a stratigraphic setting, reconstructed by manual pits dug down to the carbonate bedrock, typical of initial detachment areas (Figs 2a and 2b). In particular, the investigated sites show a total thickness for piroclastic mantle of 2.5 m (sample area 1) and 3.1 m (sample area 2), and a succession of horizons (Fig. 2b), classified with lithological and pedological criteria (Terribile et al., 2000) as well as USCS system, as follows: A) humus (Pt); B) very loose pyroclastic horizons subjected to highly pedogenetic processes with dense root apparatuses (SM); C) very loose pumiceous lapilli horizon with low degree of weathering (GW-GP); Bb) buried soil or paleosoil (SM); Cb) very loose pumiceous lapilli with low degree of weathering (GW-GP), corresponding to a deposit of a preceding eruption; Bb$_{basal}$) basal buried paleosoil (SM), corresponding to intensely pedogenised pyroclastic deposits; R) fractured carbonate bedrock with open joints filled by soil derived from the above paleosoil horizon.

Two sets of fifteen undisturbed samples, belonging to B, Bb and Bb$_{basal}$ horizons, were collected in sampling pits located from about 10 m to 20 m upstream of the main scarps: one set was devoted to laboratory engineering-geological analyses and the other to geophysical measurements. The sampling was carried out by means of a steel cylindrical sampler with a cutting edge and an inner plexiglass liner (inner diameter: 74 mm; length: 148 mm). Owing to both thinness and difficult sampling due to their extreme looseness, C horizons were not sampled in undisturbed conditions. Nevertheless, in order to increase the number of geotechnical index property determinations, other five partially disturbed samples were collected from all the above mentioned horizons. Some soil samples were collected by pushing into the soil horizon a Shelby sampler with a known inner volume, with the specific purpose to characterise index properties of C horizon (Tab. 1).

*Correspondence to: E. Piegari (esterpiegari@gmail.com ; +39 081 2538377)

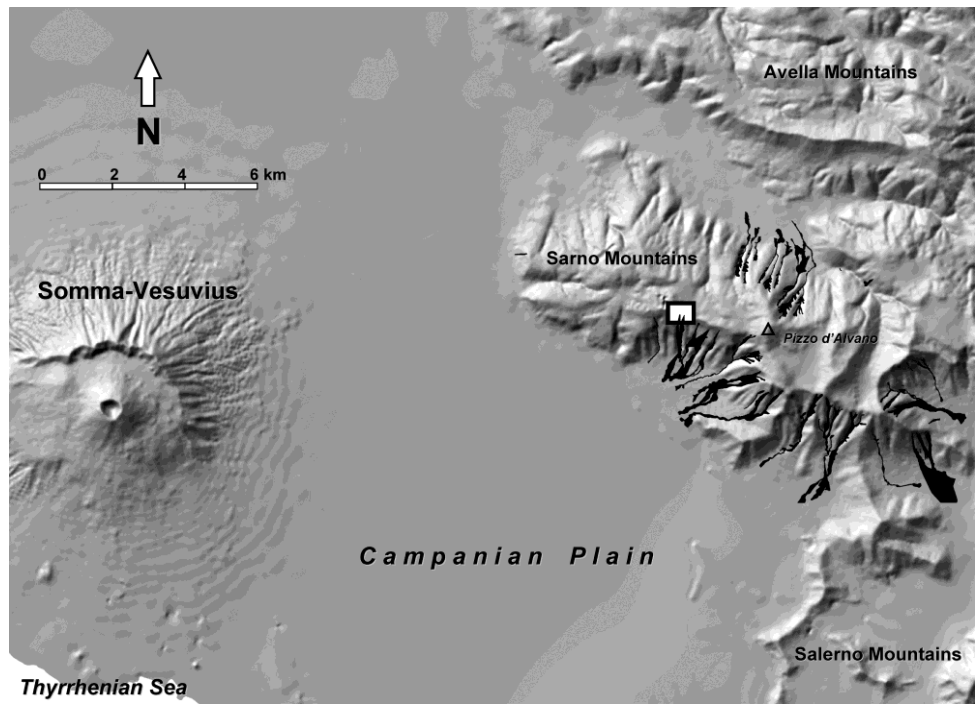

***Fig. 1****: DTM of the southern sector of the Campanian Plain that partially includes mountain ranges surrounding the Mt. Somma-Vesuvius. Black shapes correspond to the 1998's debris flows that occurred on the Sarno Mountains while the white rectangle indicates the sample areas considered in this paper.*

*Correspondence to: E. Piegari (esterpiegari@gmail.com ; +39 081 2538377)

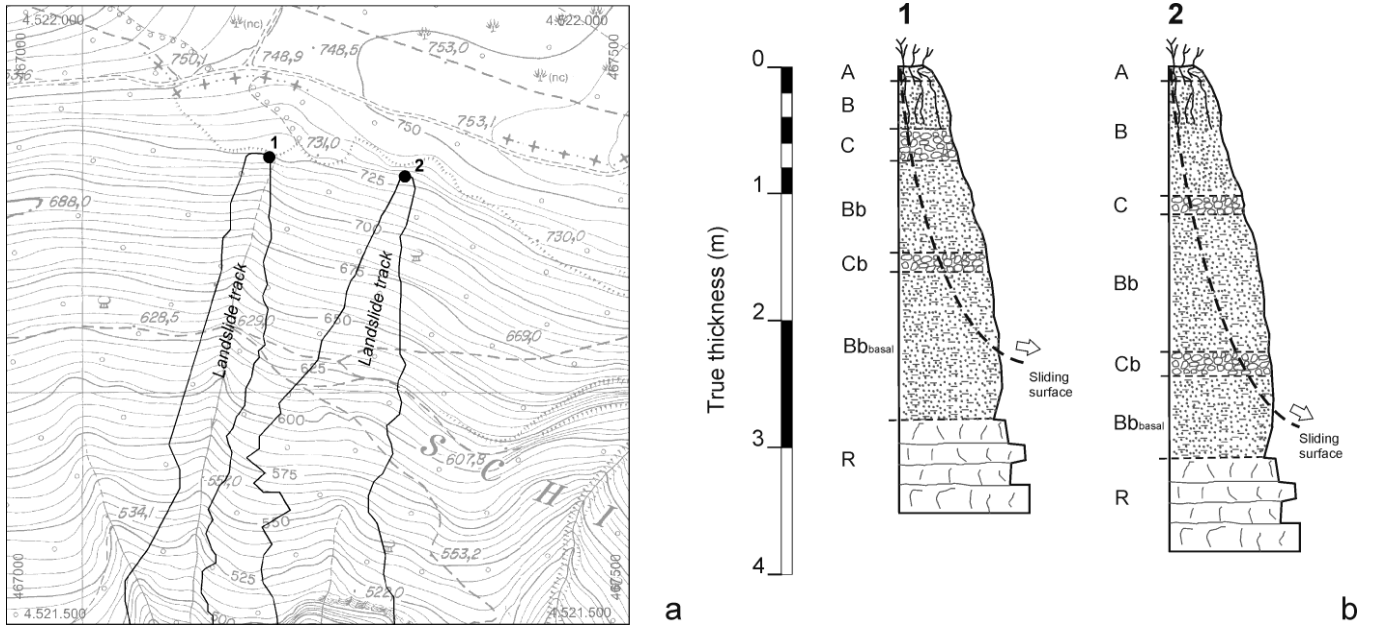

*Fig. 2*: *a) location of samplings sites (coordinates are in UTM, datum ED50); b) stratigraphic columns reconstructed in the sampling areas with the idealised indication of sliding surfaces of initial landslides.*

### 3. INDEX PROPERTIES OF ASH-FALL PYROCLASTIC SOILS

Ash-fall pyroclastic deposits are known to be soils with special engineering geological properties due to the particular microscopic structure of constituent fragments (Bell, 2000; Esposito and Guadagno, 1998). As is well known, pyroclasts (Schmidt, 1981), namely clastic materials ejected from volcanic vents whose fragmentation is due to mechanical friction or gaseous explosion during movement of lava, have a typical vescicular structure. Different types of ash-fall pyroclasts were recognised for high-explosive Plinian eruptions (hydroclastic) (Fisher and Schmincke, 1984) as glass shards, shards formed by vesciculation, pumice, pyrogenic minerals and lithic.

Pumiceous pyroclasts prevail in ash-fall pyroclastic deposits of the Mt. Somma-Vesuvius and are composed of highly vesciculated volcanic glass that, in the case of fine ash grain size (Ø < 1/16 mm), are mainly constituted of bubble-wall shards derived from broken bubbles or vescicle walls (Fisher and Schmincke, 1984). Conversely, in larger grain sizes (Fischer, 1961; Schmidt, 1981), such as coarse ash (1/16 < Ø < 2 mm) and lapilli (Ø > 2 mm), pumiceous fragments can form pyroclasts with completely or partially isolated intra-particles voids that can determine a unit weight that is lower than the water unit weight (9.807 kN/m$^3$), thus forming materials that are able to float.

Moreover, further studies, conducted on such volcanic materials (Whitam and Sparks, 1986; Esposito and Guadagno, 1998) by means of sinking tests discovered that pumiceous pyroclasts sink in water after a prolonged time due to the interconnection of intra-particle voids. This innovative result gives a different significance to the subdivision of the total porosity of these materials in both interparticle and intraparticle types. Accordingly, porosimetric analyses have shown that single pumiceous pyroclasts with diameter varying from 8.0 to 4.0 mm have a specific surface ranging, respectively, from 755.6 to 888.2 m$^2$/N (Esposito and Guadagno, 1998). Differently, the specific surface of the equivalent spherical fragment is, respectively, of $7.65 \times 10^{-2}$ m$^2$/N and $1.53 \times 10^{-1}$ m$^2$/N, if a bulk unit weight of the grains of 9.807 kN/m$^3$ is considered. Such aspect is fundamental for the comprehension of the peculiar water retention capacity of pumiceous pyroclasts that results to be greater than the other soils equivalent in grain size.

*Correspondence to: E. Piegari (esterpiegari@gmail.com ; +39 081 2538377)

To characterize index properties of the investigated pyroclastic soils, accurate geotechnical laboratory tests of the sampled soil specimens were performed, which allowed their classification with the USCS system. Laboratory analyses were carried out according to the ASTM and BS standards: ASTM D421, ASTM D2217 and ASTM D422 for grain size analyses; ASTM D4318 and BS 1377 for consistency limits.

The results of the laboratory analyses (Fig. 3 and Tab. 1) match very well with others known in the literature (Cascini et al., 2000; Crosta and Dal Negro, 2003; Guadagno and Magaldi, 2000; Bilotta et al., 2005), indicating that soils made of ash-fall pyroclastic deposits (or derived from their weathering) covering mountain slopes around the Mt. Somma-Vesuvius, present variable index properties depending on the type of pyroclasts, grain size as well as pedogenetic, erosion and transportation processes along slopes.

Among the peculiar characteristics, low values of dry unit weight ($\gamma_d$) and high values of void ratio (*e*) and porosity (*n*) can be pointed out, which reach extreme values in the case of the C horizon, made of pumiceous lapilli. Moreover, a decrease of grain size dimensions can be recognised for paleosoils (Bb horizon) in comparison with C and B horizons (Fig. 3). This decrease is particularly marked in the case of the basal paleosoil (Bb$_{basal}$) that presents the minimum grain size dimensions (Tab. 1).

*Correspondence to: E. Piegari (esterpiegari@gmail.com ; +39 081 2538377)

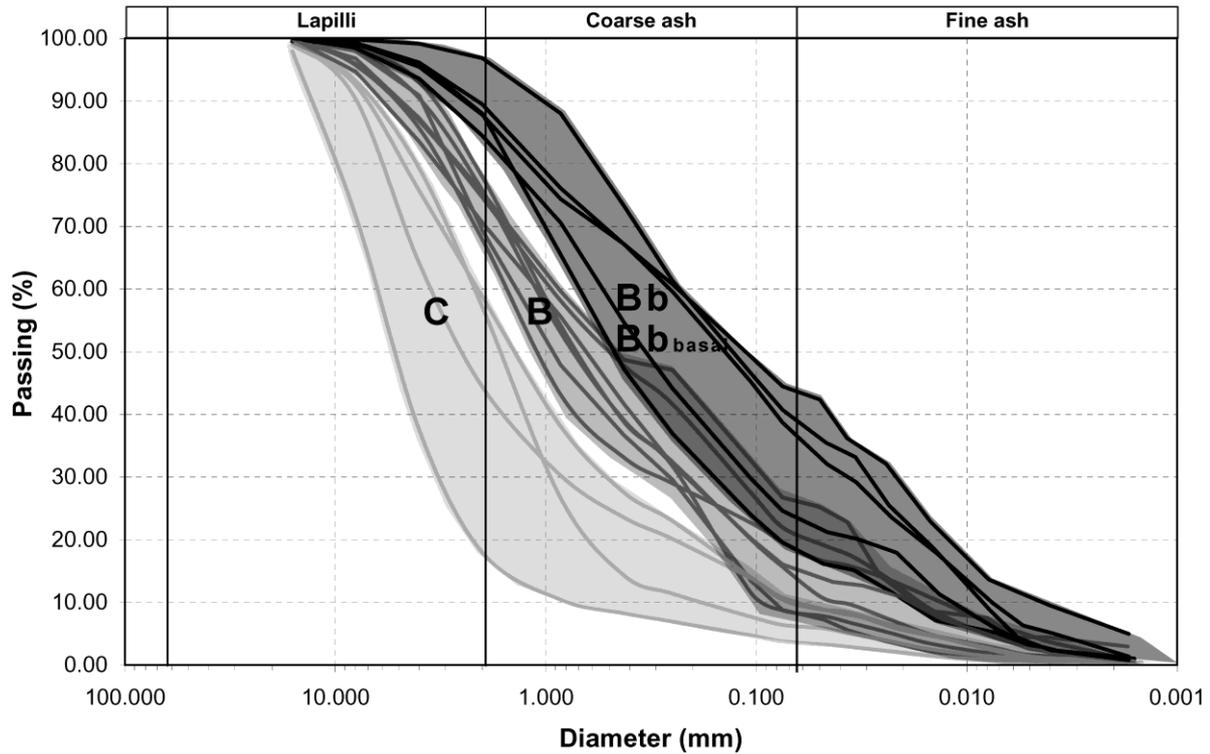

*Fig. 3*: grain size envelopes obtained for the soil horizons B, C, Bb and Bb$_{basal}$ in sample areas.

| Horizon | Description | Gs (kN/m$^3$) | $\gamma_d$ (kN/m$^3$) | e | n (%) | d$_{10}$ (mm) | d$_{60}$ (mm) | w$_L$ (%) | IP (%) | USCS |
|---|---|---|---|---|---|---|---|---|---|---|
| B | Actual pedogenised pyroclastic soil | 24.03 | 11.32 | 1.14 | 52.90 | 0.0633 | 1.040 | 39.3 | 5 < IP < 10 | SM |
| C – Cb | Unweathered pumiceous lapilli | 24.60 | 9.83 | 1.65 | 60.16 | 0.2414 | 12.63 | - | - | GW-GP |
| Bb - Bb$_{basal}$ | Buried pedogenised pyroclastic soil (paleosoils) | 24.82 | 11.17 | 1.35 | 55.24 | 0.0136 | 0.631 | 30.6 | 5 < IP < 10 | SM |

*Tab. 1*: mean values of index properties of the pyroclastic soil samples obtained for the sampled horizons.

## 4. RELATION BETWEEN WATER CONTENT AND SUCTION

The unsaturated pyroclastic soil properties characterisation was performed for samples representative of B, Bb and Bb$_{basal}$ horizons by means of the definition of the Soil Water Retention Curves (SWRCs) or soil moisture characteristic curves (Buckingam, 1904; Childs, 1969; Kutilek and Nielsen, 1994). A pressure outflow method was applied by means of the 1600 Pressure Plate Extractor, constituted of an airtight pressure vessel equipped with a porous ceramic pressure plate of 500 kPa air entry value, and an apparatus constituted of an air compressor and a pressure regulating manifold (ASTM D6836).

*Correspondence to: E. Piegari (esterpiegari@gmail.com ; +39 081 2538377)

As above mentioned, soil horizons were sampled directly in the field, in this case through direct insertion of brass cylinders (inner diameter: 53.7 mm; length: 30 mm) that, after the closing at the top and at the bottom with specific retain assemblies constituted of two ceramic porous disks, were partially soaked in free water until approximate saturation of the specimens through capillary rise occurred.

After the measurement of the saturated gross weight of the soil samples, brass cylinders and retain assemblies, the pyroclastic soil samples were put above the pressure plate in the pressure vessel. The pressure steps experimented were 0.1, 2, 5, 10, 20, 40, 90 kPa. For each of them an equilibrium between the applied air pressure and the capillary water pressure (soil suction), as well as soil water content, was recognised. The equilibrium was considered to have been reached by the observation of the annulment of water outflow coming out from the pressure vessel through the pressure plate. For each step, after the reaching of equilibrium, the gross weight of soil samples and the retain assembly were measured ($P_i$), allowing the estimation of water loss. After the last step, dry soil gross weight ($P_d$) of the samples were measured through ovendrying. Subsequently dry unit weight ($\gamma_d$) and soil volumetric water contents ($\theta_i$) were estimated for the equilibrium reached at each pressure step, adapting the basic dimensionless formula ($\theta_i = V_{wi}/V_t$ where $V_{wi}$ is the volume of water and $V_t$ the total volume) to variables of the experiment:

$$\theta_i = \frac{\dfrac{P_i - P_d}{\gamma_w}}{\dfrac{P_d - P_a}{\gamma_d}} ,$$

where $\gamma_w$ is the water unit weight and $P_a$ is the weight of brass cylinder and porous plates. The volumetric water contents at saturation ($\theta_s$) and residual ($\theta_r$), related to the last step, were also determined.

A special behaviour of the sampled pyroclastic soils, especially for those belonging to the B horizon, was observed during experimentations. Such soils suffer a significant reduction in volume up to 15% if saturated by complete soaking of soil containers. Nevertheless the adopted saturation technique based on capillary rise was verified to produce a negligible reduction of volume in soil specimens.

*Correspondence to: E. Piegari (esterpiegari@gmail.com ; +39 081 2538377)

The plotting of volumetric water content ($\theta$) vs soil matric suction ($h$), namely pore water pressure, allowed the reconstruction of SWRCs for B, Bb and Bb$_{basal}$ horizons (Fig. 4).

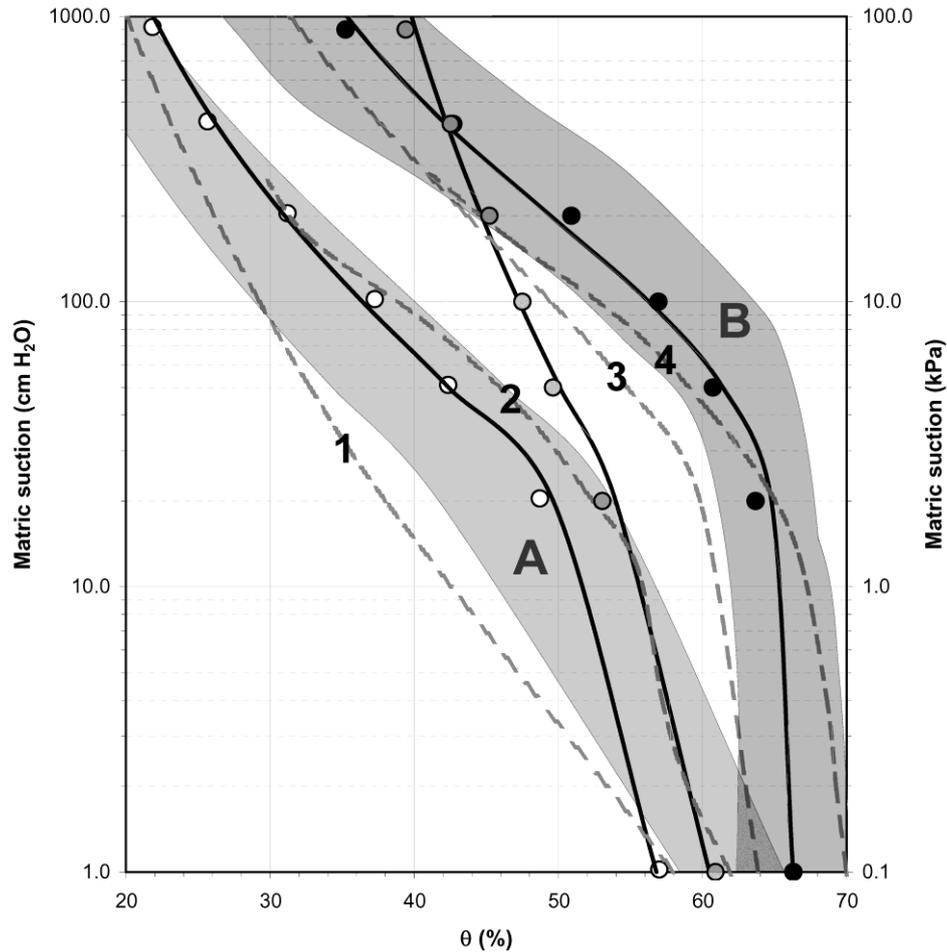

***Fig. 4***: *soil Water Retention Curves of B (white dots), Bb (black dots) and Bb$_{basal}$ (grey dots) horizons. SWRCs' envelopes are: A) B horizon (Frattini et al., 2004); B) Bb horizons (Nicotera and Papa, 2008). Single SWRCs are: 1) B horizon (Sorbino and Foresta, 2002); 2) B horizon (Terribile et al., 2000); 3) Bb horizon (Sorbino and Foresta, 2002); 4) Bb horizon (Terribile et al., 2000).*

| Van Genutchen's model parameters | $\alpha$ | n | m | $\theta s$ | $\theta r$ |
|---|---|---|---|---|---|
| B | 0.046 | 1.347 | 0.258 | 0.57 | 0.09 |
| Bb | 0.011 | 1.462 | 0.316 | 0.66 | 0.20 |
| Bb$_{basal}$ | 0.116 | 1.209 | 0.173 | 0.61 | 0.28 |

***Tab. 2***: *parameters of the van Genuchten model obtained by the optimisation procedure performed by means of RETC software on the experimental data.*

*Correspondence to: E. Piegari (esterpiegari@gmail.com) ; +39 081 2538377)

Experimental data were interpolated by means of the van Genuchten model (1980) through an optimisation calculation performed by means of the RETC software (van Genuchten, 1980; van Genuchten et al., 1994) that allowed the estimation of model parameters (α, *n* and *m* = 1-1/n) through the minimisation of the standard deviation between experimental data and the model (Tab. 2).

$$\theta(h) = \theta_r + \frac{\theta_s - \theta_r}{\left(1 + |\alpha h|^n\right)^m} .$$

Results are in good agreement with SWRCs obtained by other authors for the same soil horizons of the pyroclastic series (Terribile et al., 2000; Sorbino and Foresta, 2002; Frattini et al., 2004; Nicotera and Papa, 2008).

## 5. RELATION BETWEEN ELECTRICAL RESISTIVITY AND WATER CONTENT

Most laboratory studies on the influence of water content on electrical resistivity values were carried out especially on sandstones (Knight and Dvorkin, 1992; Taylor and Barker, 2002), carbonate and tuff rocks (Zamora et al., 1994; Carrara et al., 1994). In this paper, we describe the procedure and the results of electrical laboratory analyses on pyroclastic soils with the aim to study the behaviour of the electrical resistivity with water content variations. Specifically, resistivity measurements were performed on fifteen samples belonging to the horizons B, Bb and Bb$_{basal}$ (see §. 3.1).

### 5.1 *Experimental setting and procedure*

To simulate the in-situ saturation processes, the resistivity laboratory analyses were performed on samples saturated with raining water (electrical conductivity $\sigma = 88 \mu S/cm$) at room temperature (20 °C) and standard pressure conditions.

The measures were carried out with the four-electrode technique (Vinegar and Waxman, 1984): a known amount of stationary current is applied at the ends of the sample by means of two current electrodes and the corresponding voltage is measured by another couple of electrodes putted in the middle part of the sample (Fig. 5). We have chosen the four electrode technique to minimize the electrode polarization phenomena that can occur when a two-electrode array is applied (Roberts and Lin, 1997; Taylor and Barker, 2002). In particular, a Wenner electrode array was employed, where the four electrodes are equally spaced of a distance that, in our case (see Fig. 5), is equal to *l*; for this array the geometrical factor, *k*, of the electrode configuration is equal to $2\pi l$ (see for e.g. Reynolds, 1998).

*Correspondence to: E. Piegari (esterpiegari@gmail.com ; +39 081 2538377)

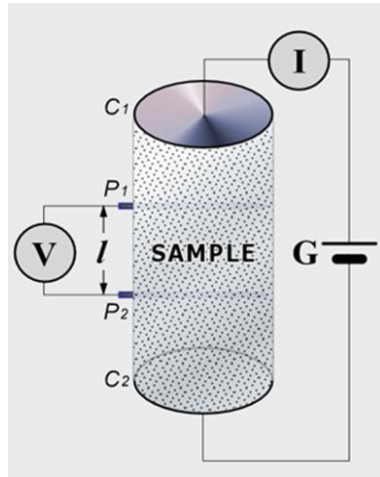

**Fig. 5**: *four electrode array: V=voltmeter, I=amperometer, l=potential electrode distance; C1,C2=current electrodes; P1,P2=potential electrodes.*

We used two very dense mesh metallic screens as current electrodes and two metallic needles as potential electrodes. The advantage of metallic screens is that they avoided the loss of material and permitted natural evaporation process. Moreover, tests performed with metallic needles and metallic screens as current electrodes showed that the resistivity values were comparable, within the range of experimental errors.

Resistivity measurements were performed with the AGI STING R1 earth resistivity meter [measure ranges: 400 KΩ to 0.1 mΩ (resistance), 0-500 full scale autoranging (voltage)], powered by a 12V battery. Beyond the earth resistivity meter and the battery, the electronic setup consisted of a precision balance ($5 \times 10^{-4}$ kg precision), a laboratory oven and a conductivity meter (Hanna Instruments HI8033) to measure the raining water conductivity used to saturate the samples.

The procedure for the resistivity measurements followed three steps. In the first step, preliminary measurements were performed to characterize the samples. In detail, we determined the total gross weights of the samples, $P_l$, the weight of the empty plexiglass tubes, $P_c$, the weights of the current electrodes, $P_e$, and the volume of the samples, $V_c$. In the second step, the samples were immerged in hydraulic heads (Fig. 6) with heights ranging from 0.035 to 0.040 m and saturated by capillary rise. Volumetric changes of the soil specimens were not appreciable. The saturation was considered completed when the water reached the upper part of the samples, which appeared translucent. We notice that such a saturation procedure might not allow the complete water filling of all interparticle voids, nevertheless this method was chosen because the alternative soaking method would have disturbed such very loose soil samples determining a relevant volume reduction especially for specimens belonging to the B horizon. We allowed the percolation of free water measuring its volume, $V_g$, and weight, $P_g$, in a graded cup. Finally, we measured the weights of the samples after the water draining, $P_{ds}$, and obtained the total weight of the saturated samples, $P_{ls}$, by means of the following expression: $P_{ls} = P_{ds} + P_g$.

*Correspondence to: E. Piegari (esterpiegari@gmail.com ; +39 081 2538377)

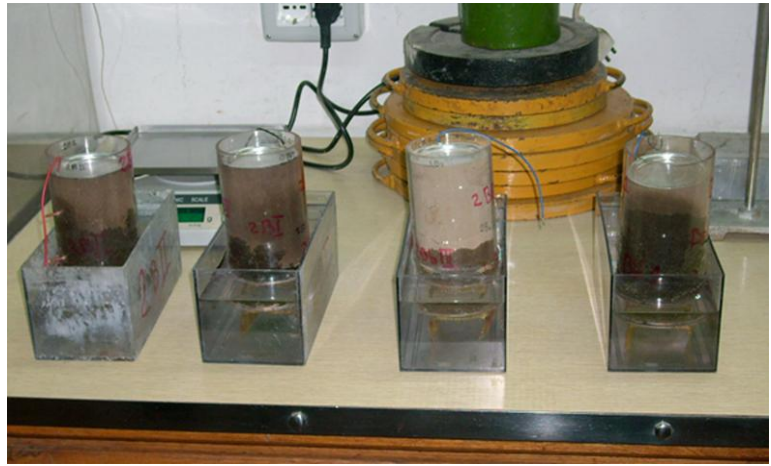

***Fig. 6****: saturation procedure of the pyroclastic samples for the resitivity measurements.*

In the third step, electrical measurements were performed at room temperature (20°C) and pressure conditions, starting from the saturated condition and continued by drying up the samples up to completely dry conditions. The decreasing levels of saturation were obtained drying the samples in oven at the temperature of 70 °C, for a period of time ranging from 30 minutes to 1 hour, during which the samples were periodically turned upside down to obtain a uniform distribution of the water content. This procedure allows us to avoid the formation of layers with different saturation degree even if initial removal of free water by means of percolation procedure prevented the layering of water content within soil samples. The measurements of the electrical resistivity, $\rho$ (= kV/I, where k=2πl, see Fig. 5), and the total sample weight, $P_l$, were performed about six hours later the kiln-dry in order to allow the thermal equilibrium with the environment at room temperature. The resistivity measurements stopped when the water content, $w$, was so small that it did not allow the passage of electrical current. The entire measurement cycle lasted for about two months. At the end of this cycle, the samples were dried up in an oven at 105 °C for 24 hours, to determine the total dry weight, $P_{ldry}$. We also calculated the values of the dry unit weight, $\gamma_d = \dfrac{P_{ldry} - (P_c + P_e)}{V_c}$, for all samples, which turned out to be comparable, within the range of experimental errors, with those obtained through geotechnical analyses (Tab. 1).

*Correspondence to: E. Piegari (esterpiegari@gmail.com ; +39 081 2538377)

*5.2 Electrical resistivity vs. water content*

At the end of the drying process, for each measurement of $\rho$ and $P_l$, we obtained the corresponding gravimetric water content, $w_i$, by the basic dimensionless formula: $w_i = P_{wi}/P_d$, where $P_{wi}$ and $P_d$ are, respectively, the water and the ovendryed sample weights (ASTM D 2216-80). Substituting in this formula the variables managed in the experiments, we used the following expression**:**

$$w_i = \frac{P_{1i} - P_{1dry}}{P_{1dry} - (P_c + P_e)}.$$

We also determined the volumetric water content, $\theta_i$, as the product between the gravimetric water content $w_i$ and the soil dry unit weight $\gamma_d$ divided by the water unit weight $\gamma_w$.

Experimental data show the increasing behaviour of $\rho$ with the decreasing of the volumetric water content, $\theta$, for the samples of the three investigated horizons (Figs. 7 ÷ 9).

Coupled electrical resistivity and water content values were experimentally determined for six samples representative of the B horizon (Fig. 7): two samples were collected at a depth of 0.70 m below the ground level (b.g.l.), while the other four samples at a depth of 0.40 m b.g.l. As it can be seen, resistivity varies from an averaged maximum value of 4000 Ωm to an averaged minimum value of 700 Ωm. The smallest values of $\rho$ are reached for volumetric water contents varying from 40% to 45%. It is worth noticing that all curves (Fig. 7) show a step in the resistivity values for $\theta \approx 10\%$, which clearly marks a different rate of change of $\rho$ for lower water contents. Such a feature could be an indication of two different types of pores: larger pores, which loose water first and smaller pores that empty later. We notice that previous analyses on the double porosity of such materials (Sorbino et al., 2006) would find a confirmation in our electrical characterisation of their properties.

*Correspondence to: E. Piegari (esterpiegari@gmail.com ; +39 081 2538377)

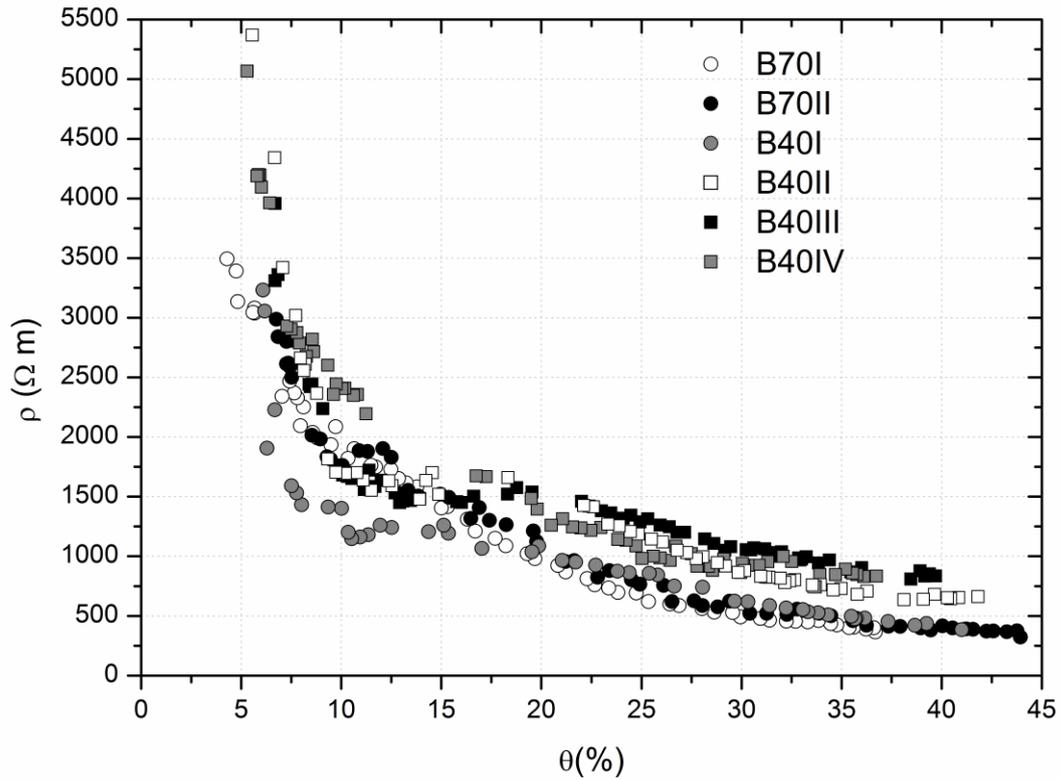

**Fig. 7**: *electrical resistivity of the B samples as a function of the volumetric water content.*

Fig. 8 shows the electrical resistivity and water content relationship, $\rho$ vs. $\theta$, for the six investigated samples representative of the Bb horizon (Fig. 8): the circles refer to samples collected at a depth of 1.40 m b.g.l., while the squares represent the samples collected at a depth of 2.00 m b.g.l. As it can be observed, $\rho$ varies from an averaged maximum value of 6000 Ωm to an averaged minimum value of 400 Ωm. The smallest values of $\rho$ are reached for volumetric water contents varying from 60% to 65%. Unlike the data relative to the B horizon, the behaviour of $\rho$ of the Bb samples does not seem to show a discontinuity for $\theta \approx 10\%$ even if, for larger $\theta$, the result of a power-law fit indicates that for both the horizon B and Bb the experimental data seem to obey to the Archie law (Archie, 1942) with a negative exponent of the order of 1.

Finally, Fig. 9 displays the behaviour of $\rho$ as a function of $\theta$ for the three investigated samples of the lowest horizon, Bb$_{basal}$. The curves show a discontinuity for volumetric water contents around 10% and reach the smallest values of $\rho$ for volumetric water contents varying from 50% to 60%. For $\theta > 15\%$ the result of a power-law fit shows a negative exponent larger than 1. Such a feature could reflect the different composition of such samples.

*Correspondence to: E. Piegari (esterpiegari@gmail.com ; +39 081 2538377)

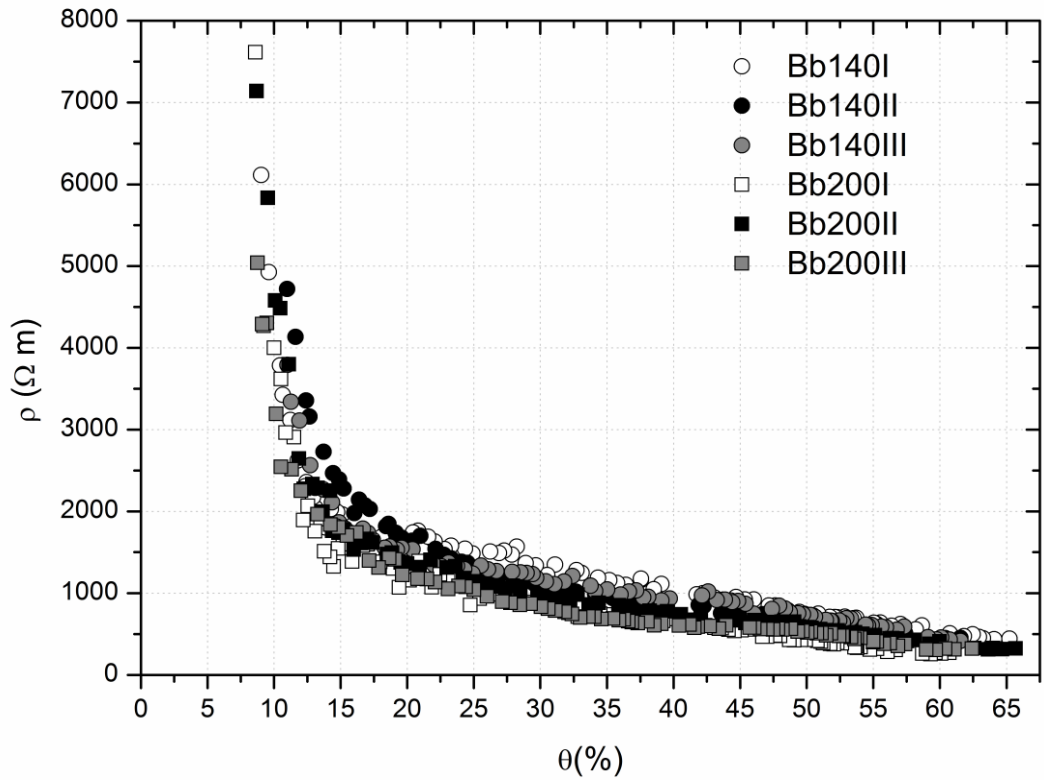

*Fig. 8*: electrical resistivity of the Bb samples as a function of the volumetric water content.

*Correspondence to: E. Piegari (esterpiegari@gmail.com ; +39 081 2538377)

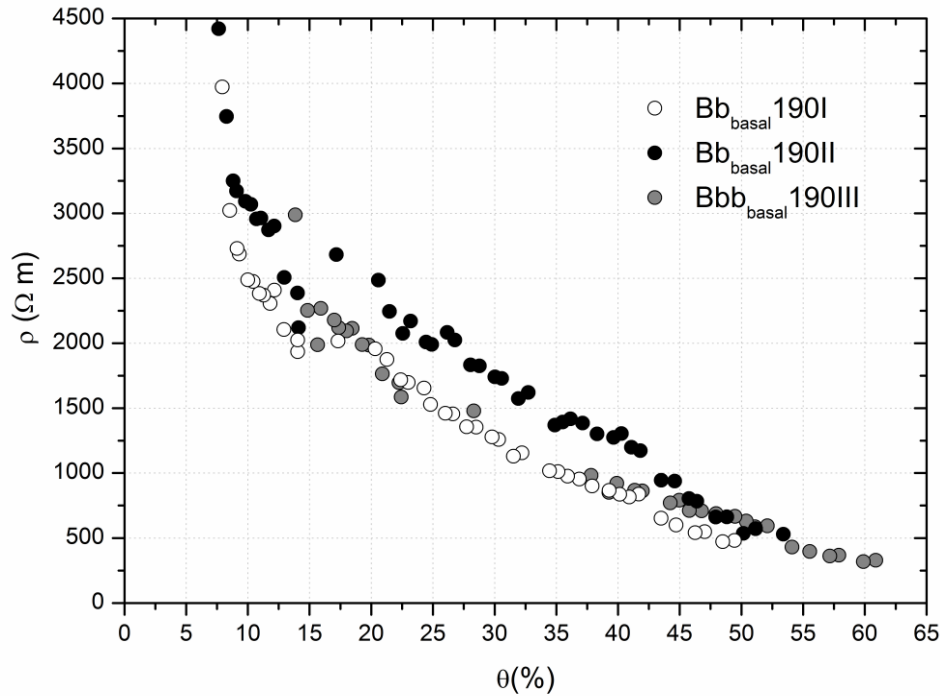

***Fig. 9***: *electrical resistivity of the Bb$_{basal}$ samples as a function of the volumetric water content.*

6.  **RELATIONSHIP BETWEEN ELECTRICAL RESISTIVITY AND SOIL MATRIC-SUCTION**

As it is well-known, slope stability is affected by several factors, such as pore water pressure, soil strength parameters, cohesion and slope angle. In particular, in the case of steep slopes, where inclination is larger than internal friction angle, suction values are crucial since stability is essentially controlled by the apparent cohesion due to suction under unsaturated conditions (Fredlund et al., 1978; Lu and Likos, 2006). Such conditions are very common for the investigated slopes (Olivares and Picarelli, 2003) and, therefore, the determination of SWRC is of great interest. Particularly, for the triggering of rainfall-induced landslide events the recognition of small values of the suction is fundamental as suction vanishes with increasing water content (see Fig. 4). On the other hand, electrical resistivity also strongly depends on water content of the soil, by showing a power-law decrease with the water content (see Figs. 7÷ 9). Therefore, a possible correlation between electric resistivity and soil matric suction may be investigated in the attempt to better characterise the region close to the saturated condition that is a critical state for the stability of pyroclastic covers.

*Correspondence to: E. Piegari (esterpiegari@gmail.com ; +39 081 2538377)

The correlation between electrical resistivity and soil matric suction has been investigated on three samples representative of B, Bb and Bb$_{basal}$ horizons, considering, respectively, the continuous function derived from empirical correlations (Figs. 7, 8 and 9) and from the van Genuchten models (Fig. 4 and Tab. 2). Graphical correlations (Fig. 10) show complex nonlinear relationships that are in good agreement with the physical characteristics of the pyroclastic soil samples. Among them it is possible to point out the marked divergence in soil suction for Bb and Bb$_{basal}$ horizons in comparison with the B horizon for electrical resistivity values greater than 600 Ωm, which accounts for difference in grain size characteristics and, therefore, in water retention properties.

In order to give a practical application of the outlined correlations, experimental data were interpolated through polynomial regressions (Fig. 10).

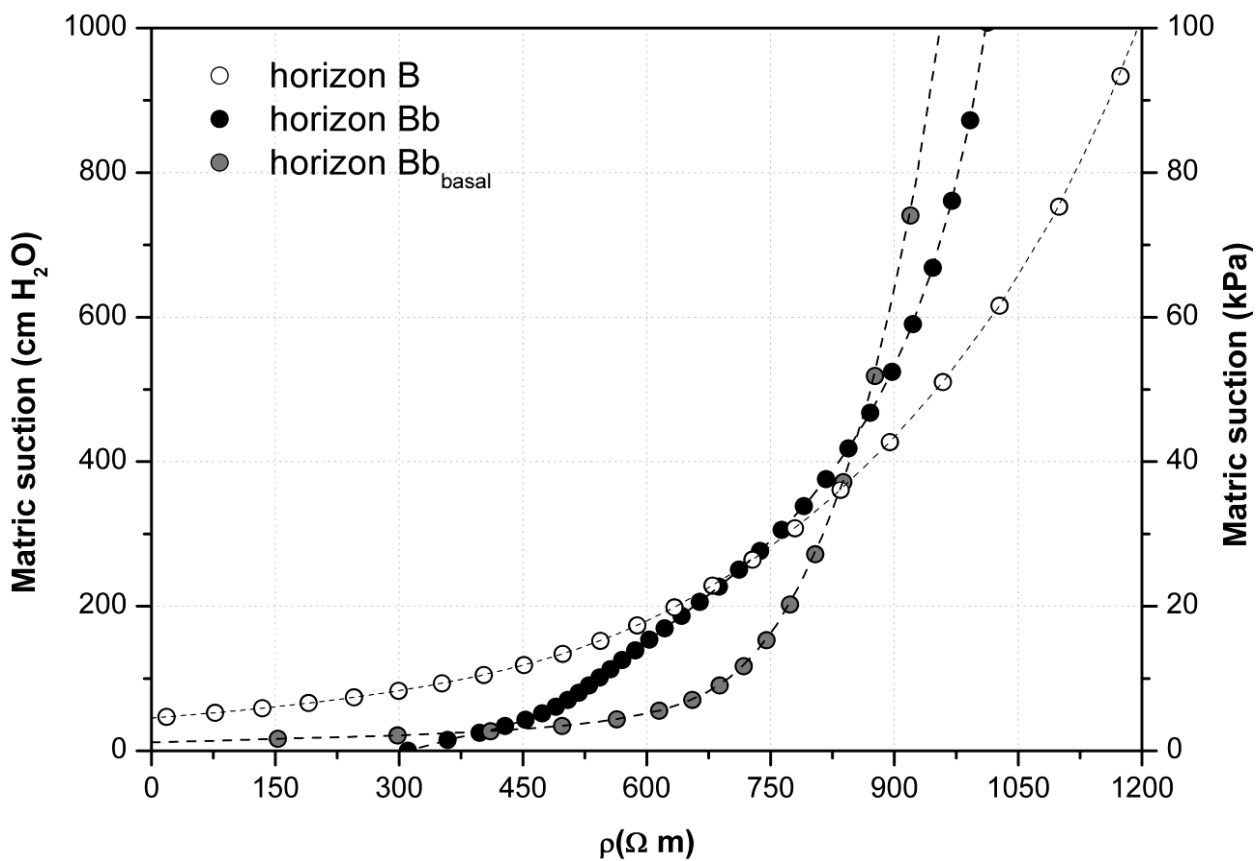

*Fig. 10*: *soil matric suction as a function of the electrical resistivity.*

We notice that for soil matric suction values smaller than 10 kPa the curves representative of the shallowest (B) and the deepest (Bb$_{basal}$) horizons show the largest variations in resistivity, which starts to vanish from values of about 300 Ωm for the B horizon and 700 Ωm for the Bb$_{basal}$ horizon. On the other hand, the correlation between electrical resistivity and soil matric suction seems to appear different for the intermediate horizon Bb, for which electrical resistivity values corresponding to vanishing suction stay finite and at about 300 Ωm. It is also worth noticing that for very small suction values, the relation between resistivity and soil suction can be approximated by a straight line.

*Correspondence to: E. Piegari (esterpiegari@gmail.com ; +39 081 2538377)

Actually, the Factor of Safety (FS) for stability analysis of shallow landslides, is straightforwardly considered dependent also on the increase in shear strength due to soil matric suction (apparent cohesion), which can not be neglected in the variable saturation conditions existing in the vadose zone during a rainstorm (Godt et al., 2009; Lu and Likos, 2006). Thus, the exposed results might be of great relevance for the landslide hazard assessment of pyroclastic covers because the discovered empirical relationships between electrical resistivity and soil matric suction could be directly implemented in stability calculations. Moreover, due to the high sensitivity of the electrical resistivity close to saturation (h < 10 kPa), which accounts for well measurable strong variations, this parameter can be considered very useful to recognise the soil matric suction in a volume representative for shallow landslide initiation.

Experimental results corroborate the dependence between FS and electrical resistivity recently introduced by two of us as a first attempt to relate resistivity to stability (Piegari et al., 2009a).

## 7. CONCLUSIONS

In the last fifteen years, the great number of shallow landslides that occurred in the mountain areas of the Campania region that involved pyroclastic covers have pointed out the need for a detailed characterisation of such materials. In particular, loose ash-fall pyroclastic soils often mantle steep slopes with inclination comparable with soil internal friction angles, and, therefore, slope stability is mainly controlled by the component of shear strength due to cohesion, to which soil suction also gives a not negligible contribute in combination with the strength of root apparatuses. Since critical rainfalls are a major trigger for landslides involving ash-fall pyroclastic soils, an accurate characterization of the behaviour of parameters crucial for slope stability is needed in presence of large water contents. With this aim in mind, we carried out engineering-geological and geophysical laboratory analyses on unsaturated pyroclastic soil samples collected on Sarno Mountains (southern Italy) to better characterise the conditions close to saturation, which are critical for stability of shallow landslides.

Analyses of geotechnical and hydraulic unsaturated properties confirmed the peculiarity of pyroclastic soils, already demonstrated by previous papers. In particular, Soil Water Retention Curves were obtained for samples belonging to the three investigated soil horizons.

*Correspondence to: E. Piegari (esterpiegari@gmail.com ; +39 081 2538377)

At the same time, geophysical analyses provided the determination of the first characteristic curves electrical resistivity vs. volumetric water content. In particular, an experimental procedure for finding the values of the electrical resistivity of pyroclastic samples at different water contents has been illustrated.

As both matric suction and electrical resistivity strongly depend on water content, we have determined an empirical relationship between such variables for each investigated horizon. Since infiltration process during the rainy season produce a progressive vanishing of the matric suction up to values recorded in the late Winter that commonly ranging between 5 kPa to 15 kPa (Sorbino and Foresta, 2002; Sorbino, 2005), which can be transitorily lowered to positive pore pressure (saturation) by the occurrence of heavy rainstorms and then determining landslide triggering, we look for electrical resistivity values corresponding to small suction. We find that for soil matric suction values smaller than 10 kPa the curves representative of the shallowest (B) and the deepest ($Bb_{basal}$) horizons show the largest variations in resistivity, which starts to vanish from values of about 300 $\Omega$m for the B horizon and 700 $\Omega$m for the $Bb_{basal}$. Furthermore, from the analyses, it results that even if electrical resistivity and matric suction are related by a nonlinear function, for soil conditions close to the saturated state small values of resistivity appear directly related to small suctions. This result turns out to be of great relevance for the hazard assessment of pyroclastic covers, as it allows to directly relate resistivity to FS through the implementation of discovered empirical relationship in limit equilibrium slope stability calculations.

In conclusion, the results presented in this paper demonstrated that an integration of engineering-geological and geophysical methods can provide useful information for a better understanding of the pyroclastic soil response to high water contents responsible of slope failure. In particular, our analysis reveals the existence of a possible link between the geotechnical and geophysical approaches for slope stability analysis. The first based on measurements of mechanical soil properties, like friction angles, cohesion, suction, etc.., performed on very small soil volumes, the second one based on measurements of physical quantities, like resistivity, on soil volumes much more representative for shallow landslide triggering. Indeed, field measurements of electrical resistivity by means of tomographic technique (Di Maio et al., 2007; Piegari et al., 2009a) allow to investigate volumes of soils larger than those controlled by tensiometers.

Moreover, the discovered empirical relationships can be considered useful and cost-effective tools for directly monitoring the hydrological state of pyroclastic covers thus to identify threshold values of resistivity/suction on which to set up an early warning system. In such a meaning, a continuous monitoring of electrical resistivity could constitute an early warning system allowing the recognition of conditions differently prone to landsliding, such as those usually occurring at the end of the rainy season (5-15 kPa) or during heavy rainstorms (0-5 kPa), especially if occurring in the rainy season (Sorbino, 2005). Indeed, such system could be more reliable of empirical hydrological thresholds based only on actual and antecedent rainfalls (Crozier and Eyles, 1980; Chirico et al., 2000; De Vita, 2000) that do not account for the real hydrological conditions of soil mantle depending on complex hydrological processes, like evapotranspiration and seepage.

*Correspondence to: E. Piegari (esterpiegari@gmail.com ; +39 081 2538377)

6. ACKNOWLEDGEMENTS

We thank Dr Nicola Roberti and Dr Enrico Di Clemente, respectively, researcher of Applied Geophysics and technician in chief of the Engineering Geology and Geotechnics Laboratory at the Department of Earth Sciences of the University of Naples "Federico II", for the assistance during the conduction of soil testing and geophysical experimentations. We also thank the Reviewers for their comments that help us to improve the manuscript. We acknowledge financial support from the Ministry for Education, University and Research (MIUR-Italy): PRIN Project (2007) "*Analysis and assessment of susceptibility and hazard triggered by extreme natural events (precipitations and earthquakes)*".

*Correspondence to: E. Piegari (esterpiegari@gmail.com) ; +39 081 2538377)

*Correspondence to: E. Piegari (esterpiegari@gmail.com ; +39 081 2538377)

*Correspondence to: E. Piegari (esterpiegari@gmail.com ; +39 081 2538377)

*Correspondence to: E. Piegari (esterpiegari@gmail.com ; +39 081 2538377)